\documentstyle[12pt]{article}
\begin{document}

\baselineskip=22.5pt

\begin{flushright}
{\bf LMU-07/94}\\
{\bf PVAMU-HEP-94-5}\\
{June 1994}
\end{flushright}

\vspace{0.5cm}

\begin{center}
{\Large\bf On the Unitarity Triangles of the CKM Matrix}
\vskip .5in

{\bf Dan-di WU\footnote{Supported in part by the U.S.
 Department of Energy under contract DE-FG05-92ER4072S}}\\
{\sl HEP, Box 355, Prairie View A\&M University, Prairie View, TX 77446, USA}
\vskip .21in

and
\vskip .21in

{\bf Zhi-zhong XING\footnote{Alexander-von-Humboldt Research Fellow}}\\
{\sl Sektion Physik der Universit${\ddot a}$t M${\ddot u}$nchen, D-80333
Munich, Germany}
\end{center}

\vskip 1.in

\begin{abstract}

The unitarity triangles of the $3\times 3$ Cabibbo-Kobayashi-Maskawa (CKM)
matrix are studied in a systematic way.
We show that the phases of the nine CKM rephasing invariants
are indeed the outer angles of the six unitarity triangles
and measurable in the $CP$-violating decay modes of $B_{d}$ and $B_{s}$ mesons.
An economical notation system is introduced for describing properties of
the unitarity triangles. To test unitarity of the CKM matrix
we present some approximate but useful relations among the sides and angles
of the unitarity triangles, which can be confronted with the accessible
experiments
of quark mixing and $CP$ violation.

\end{abstract}

\newpage

	In the standard electroweak model,  the $3\times 3$ Cabibbo-Kobayashi-Maskawa
(CKM) matrix $V$ describes flavor mixing and $CP$ violation [1]. Unitarity is
the only
constraint, imposed by the model itself, upon $V$. Deviations from the minimal
standard model may give rise to or fake as unitarity breaking. For example,
unitarity of the $3\times 3$ CKM matrix will not hold
if there are fourth-family quarks or exotic charge $-1/3$ quarks which mix with
the standard
down-type quarks. It is therefore very important to stringently test unitarity
of the
CKM matrix. Recently, a lot
of theoretical and experimental attention has been paid to this subject [2-8],
in particular,
to the six unitarity triangles. In Ref. [4], Aleksan et al have proved that
four independent inner angles of the six triangles can provide a complete
parametrization of the
CKM matrix. Their results sharpen the importance of measuring $CP$ violation in
$B$-meson decays
so as to determine the entire quark mixing matrix $V$.

	In this paper we shall make a systematic study of the unitarity triangles
using
the rephasing invariant quantities of the CKM matrix defined previously by one
of the authors [9].
We show that the phases of the nine rephasing invariants are indeed the outer
angles of the six unitarity triangles and directly related to the
$CP$-violating
asymmetries in some distinct decay modes of $B_{d}$ and $B_{s}$ mesons.
To fully test unitarity of the standard model we give some useful
relations among the sides and angles of the unitarity triangles, which
can be confronted with the forthcoming experiments of quark mixing and $CP$
violation.
It should be emphasized that our starting point of view differs from that in
Ref. [4], nevertheless,
some results of these two papers are in agreement with and supplementary to
each other.
In addition, we give a more economical notation scheme for the six unitarity
triangles and
present a practical way to check unitarity of the CKM matrix with the
accessible experimental data.
Recognizing that a few distinct and concise
parametrization forms of the CKM matrix have been extensively applied to the
studies of flavor mixing and $CP$ violation [6-8], the work done here is
expected to provide a systematic approach to the same problem.

	The unitarity triangles are defined by the following orthogonality relations:
\begin{equation}
\begin{array}{lll}
\displaystyle\sum_{\alpha =d,s,b}V_{i\alpha}V^{*}_{j\alpha} & = &0 \; ,
\;\;\;\;\;\; (i < j)\; ;\\
\displaystyle\sum_{i=u,c,t}V_{i\alpha}V^{*}_{i\beta} & = & 0 \; ,
\;\;\;\;\;\; (\alpha < \beta) \; .
\end{array}
%		(1)
\end{equation}
Here and hereafter, Latin subscripts run over the up-type quarks $u,c$, and
$t$; while
Greek ones run over the down-type quarks $d,s$, and $b$.
It is known that there exist nine rephasing invariants for the $3\times 3$
CKM matrix $V$ [9]:
\begin{equation}
\Delta_{i\alpha}\; \equiv \;
V_{j\beta}V_{k\gamma}V^{*}_{j\gamma}V^{*}_{k\beta}\; ,
\;\;\;\;\;\;\;\;\; (i,j,k \; {\rm and}\; \alpha,\beta,\gamma \; {\rm co}-{\rm
cyclic})\; .
\end{equation}
%		(2)
Here each $\Delta_{i\alpha}$ is a product of the four matrix elements in the
$2\times 2$ submatrix without
$V_{i\alpha}$. It is crucial for such products that  either two diagonal or
skew diagonal
elements are complex conjugated. The co-cyclic rule of the subscripts ensures a
definite set of
$\Delta_{i\alpha}$.
In the subsequent discussions we shall imply co-cyclic
permutation of the subscripts
whenever three different Latin and (or) Greek subscripts simultaneously appear
in a formula.
Note that the phases of $\Delta_{i\alpha}$, denoted by
\begin{equation}
\Phi_{i\alpha}\; \equiv \; \arg \Delta_{i\alpha} \; , \hskip 1.in (-\pi
<\Phi_{i\alpha} \le \pi) \; ,
\end{equation}
%(3)
are also invariant under quark phase redefinitions.
In the complex plane,
$\Phi_{i\alpha}$ is not only the relative angle of the two vectors
$V_{j\beta}V^{*}_{j\gamma}$ and $V_{k\beta}V^{*}_{k\gamma}$ but also
that  of another two vectors
$V_{j\beta}V^{*}_{k\beta}$ and  $V_{j\gamma} V^{*}_{k\gamma}$.
Thus $\Phi_{i\alpha}$ is indeed an outer angle shared by two different
unitarity triangles (see Fig. 1).
Unitarity of the $3\times 3$ CKM matrix
requires that all the nine quantities $\Delta_{i\alpha}$ have
a common imaginary part [10], defined as
\begin{equation}
J \; \equiv \; {\rm Im}\, \Delta_{i\alpha}\;=\; |\Delta_{i\alpha}|\sin
\Phi_{i\alpha} \; ,
%		(4)
\end{equation}
which is a measure of $CP$ violation in the standard model. From Eqs. (3) and
(4) one can observe that
all  nine $\Phi_{i\alpha}$ should have the same sign as $J$.

With Eqs. (2) and (3), it is straightforward to obtain the sum rules for
$\Phi_{i\alpha}$, which are similar to the orthogonality relations given in Eq.
(1):
\begin{equation}
\begin{array}{ccc}
\displaystyle\sum_{i=u,c,t}\Phi_{i\alpha} & = & \pm 2\pi \; ,\;\;\;\;\;\;\;\;\;
(\alpha = d,s, \; {\rm or}\; b)\; ;\\
\displaystyle\sum_{\alpha =d,s,b}\Phi_{i\alpha} & = & \pm 2\pi \; ,
\;\;\;\;\;\;\;\;\; (i=u,c, \; {\rm or}\; t)\; ,
\end{array}
%		(5)
\end{equation}
where the plus (minus) sign corresponds to
 $J>0$ ($J<0$). Since
phase redefinitions of the quark fields  cannot change the sign of $J$, the
sign of $\Phi_{i\alpha}$
is definitely part of the information on the CKM unitarity.
The current data on $|V_{ub}/V_{cb}|$,
$x_{d}$ ($B^{0}_{d}-\bar{B}^{0}_{d}$
mixing), and $\epsilon_{K}^{~}$
(the $CP$-violating parameter in the kaon system) already
indicate a positive $J$ [11]. This implies $\Phi_{i\alpha}> 0$, corresponding
to
the anti-clockwise triangles in the complex plane (See Fig. 1).
Subsequently we shall carry out the analysis under the condition of $J > 0$.
It is easy to follow a parallel analysis for the case of $J < 0$.

It proves convenient to use the capital form of those indices in
the parentheses of Eq. (5) to name the unitarity triangles. For example, the
${\it D}$
triangle has three sides $S_{uD}\equiv |V_{us}V^{*}_{ub}|$,
$S_{cD}\equiv |V_{cs}V^{*}_{cb}|$,
and $S_{tD}\equiv |V_{ts}V^{*}_{tb}|$, which face  three outer
angles $\Phi_{ud},\,\Phi_{cd}\,$, and $\Phi_{td}$, respectively.
Here the rule is that the CKM matrix elements with the subscript $d$
never appear in the $D$ triangle. In other words,
the $D$ triangle can only be formed by the elements in the second and third
columns of the CKM matrix.
It should be noted that the six unitarity triangles have only
nine different outer angles ($\Phi_{i\alpha}$), although  they have eighteen
different sides ($|V_{j\beta}V^{*}_{j\gamma}|$ and
$|V_{j\beta}V^{*}_{k\beta}|$).

The inner angles of the unitarity triangles,
denoted by\footnote{Analytically, $\omega_{i\alpha}\equiv |
\arg(-\Delta^*_{i\alpha}) | $.}
$\omega_{i\alpha}$, satisfy the sum rules of the Euclidean geometry:
\begin{equation}
\displaystyle\sum_{i=u,c,t} \omega_{i\alpha}=\displaystyle \sum_{\alpha =d,s,b}
\omega_{i\alpha}=\pi \; .
\end{equation}
%(6)
They are related to the outer angles $\Phi_{i\alpha}$ through
\begin{equation}
\omega_{i\alpha}\; =\; \pi \mp \Phi_{i\alpha} \; ,
\end{equation}
%(7)
where the minus (plus) sign corresponds to $J>0$ ($J<0$). From the above
relations
we see that the roles of outer and inner angles
are equivalent in discussing flavor mixing and $CP$ violation. For example, one
can use
$\Phi_{us}$, $\Phi_{cs}$, and $\Phi_{ts}$ to describe the $S$ triangle, whose
inner angles are
commonly defined by $\alpha$, $\beta$, and $\gamma$ in the literature [12-14].

As to the magnitudes of the outer angles $\Phi_{i\alpha}$,
Eq. (5) shows that the smaller $|\Delta_{i\alpha}|$, the larger
$\sin\Phi_{i\alpha}$.
In particular, we have
\begin{equation}
\sin \Phi_{tb}\; \sim \;\lambda^{4}\; , \;\;\;\;\;\;\;\; \sin \Phi_{ud}\; \sim
\; \lambda^{2}\; ,
%		(8)
\end{equation}
where $\lambda^{2}\sim \frac{1}{20}$.  The magnitudes of the remaining seven
$\sin \Phi_{i\alpha}$ are of order $O(1)$.

It should be noted that one of the six relations in Eq. (5) is trivial.
 This implies that only four of the nine outer angles $\Phi_{i\alpha}$ are
independent.
It is easy to prove that such four independent $\Phi_{i\alpha}$ are enough to
fully
determine the magnitudes of the
CKM matrix elements $|V_{i\alpha}|$. Indeed, we have the following
relations between the sides and outer angles of the unitarity triangle $K$ (or
$\Gamma$):
\begin{equation}
\begin{array}{ccc}
S_{K\alpha}\sin\Phi_{k\beta} & = & S_{K\beta}\sin \Phi_{k\alpha} \; , \\
S_{i\Gamma}\sin\Phi_{j\gamma} & = & S_{j\Gamma}\sin \Phi_{i\gamma} \; .
\end{array}
%		(9)
\end{equation}
As a result, all nine $|V_{i\alpha}|$ can be given in terms of
$\Phi_{i\alpha}$,
for example [15], from
\begin{equation}
\frac{|V_{i\alpha}|^{2}}{|V_{i\beta}|^{2}} \; =\;
\frac{S_{K\alpha}S_{I\beta}S_{J\alpha}}
{S_{K\beta}S_{I\alpha}S_{J\beta}}
\; =\; \frac{\sin \Phi_{k\alpha} \sin \Phi_{i\beta} \sin \Phi_{j\alpha}}
{\sin \Phi_{k\beta} \sin \Phi_{i\alpha} \sin \Phi_{j\beta}} \;
%		(10)
\end{equation}
and the normalization condition $\displaystyle\sum_{i
=u,c,t}|V_{i\alpha}|^{2}=1$.
At this point, it is also worthwhile to give the relations between the sides
and
the cosines of the outer angles:
\begin{eqnarray}
\cos\Phi_{k\gamma} \; = \; \displaystyle
{S_{K\gamma}^2-S_{K\alpha}^2-S_{K\beta}^2\over 2S_{K\alpha}S_{K\beta}}
\; =\; \displaystyle {S_{k\Gamma}^2-S_{i\Gamma}^2-S_{j\Gamma}^2 \over
2S_{i\Gamma}S_{j\Gamma}} \; .
\end{eqnarray}
%(11)
The cosines should be more sensitive to small alterations of the angles if
their values are close to $\pi /2$.

	In practice, it will be convenient to choose $\Phi_{us}, \Phi_{cs},
\Phi_{cd}$,
and $\Phi_{ud}$ as a set of four independent parameters. These four phases are
directly
related to the observables of $CP$ violation in neutral $B$ decays [16], and
therefore
can be well determined in the near future. For illustration, we list in Table 1
some promising decay modes of $B_{d}$ and $B_{s}$ mesons, which can be used to
probe the outer angles
$\Phi_{i\alpha}$ in experiments.

	We proceed with some discussions about testing unitarity of the CKM matrix.
Due to various experimental limitations and theoretical uncertainties, a
perfect test of the CKM unitarity
is in reality impossible. Hence a practical
or approximate check of the unitarity conditions is worth pursuing. According
to Eq. (8),
\begin{equation}
\Phi_{tb}=\Phi_{ud}=\pi\;
%		(12)
\end{equation}
is a good approximation up to $O(\lambda^{2})$. Using Eq. (12) we obtain
four simple sum rules for the outer angles $\Phi_{i\alpha}$:
\begin{equation}
\begin{array}{ccc}
\Phi_{td}+\Phi_{ts}&=&\pi +O(\lambda^{4}) \; ,\\
\Phi_{ub}+\Phi_{cb}&=&\pi +O(\lambda^{4}) \; , \\
\Phi_{us}+\Phi_{ub}&=&\pi +O(\lambda^{2})\; ,\\
\Phi_{cd}+\Phi_{td}&=&\pi +O(\lambda^{2})\; .
\end{array}
\end{equation}
%		(13)
In a slightly different form, the above relations are expressed as
\begin{eqnarray}
\begin{array}{ccccc}
\sin\Phi_{td}&=&\sin\Phi_{ts}&=&\sin\Phi_{cd} \; ,\\
\sin\Phi_{ub}&=&\sin\Phi_{cb}&=&\sin\Phi_{us} \; ,
\end{array}
\end{eqnarray}
%		(14)
which are valid up to $O(\lambda^2)$. From Eq. (12), some similar approximate
relations may be obtained,
up to $O(\lambda^{4})$, for the sides of the unitarity triangles:
\begin{eqnarray}
\begin{array}{ccc}
S_{{\it U}b}=S_{{\it U}s} \;, & \;\;\;\;\;&  S_{c{\it D}}=S_{t{\it D}} \; ;\\
S_{{\it T}d}=S_{{\it T}s} \;, &            & S_{c{\it B}}=S_{u{\it B}} \; .
\end{array}
\end{eqnarray}
%		(15)
Clearly, all these approximate relations are experimentally accessible in the
near future.
Any evidence which conflicts with the above relations should be a signal of
unitarity breaking.

For a complete test of unitarity, one has to check at least
nine independent relations, in which at least one normalization condition is
included and
at least one outer (or inner) angle is involved.
Fortunately, the six elements of the first two rows in the CKM matrix have
already been measured and they
satisfy the normalization conditions up to $O(\lambda^4)$ [5].
Except for $S_{Ub},S_{Us}$, and $S_{tD}$, the other sides in Eq. (15) can now
be determined very well.
The relations $S_{Td}=S_{Ts}$ and $S_{cB}=S_{uB}$ are consistent with
the relevant data. No doubt, a precise determination of
$|V_{ub}|$ is possible in further experiments of $B$-meson decays [17]. On the
other hand, measurements
of the top-quark lifetime will sharpen the value of $|V_{tb}|$, which is
expected to be unity under
the unitarity conditions [5]. The magnitudes of $V_{td}$ and $V_{ts}$ are
considerably difficult to be
measured in direct decays of the top quark [18]. They may be determined
indirectly from the
forthcoming precise data on $B^{0}_{d}-\bar{B}^{0}_{d}$ and
$B^{0}_{s}-\bar{B}^{0}_{s}$ mixings,
respectively.

The problem is that one has not obtained any  definite information on
the angles of the unitarity triangles to check the relations in Eqs. (12) and
(14).
Progress is therefore dependent upon measuring $CP$
violation in nonleptonic $B$ decays, as illustrated in Table 1. In the
literature, many approaches have been proposed to reliably extract the CKM
phase from
the expected large $CP$ asymmetries in the $B$-meson system [12-14, 16]. All
these discussions are
useful to determine the phases $\Phi_{i\alpha}$.

	In summary, we have explored various relations among the outer
and inner angles as well as the sides
of the CKM unitarity triangles in a rephasing invariant way. The outer angles
$\Phi_{i\alpha}$
are emphasized in describing flavor mixing and $CP$ violation within the
standard model.
To test unitarity of the $3\times 3$ CKM matrix, we have presented some
approximate
but useful relations among the outer angles and sides, which
are accessible to experiments in the near future. \\

	One of us (Z.Z.X.) would like to thank Professor H. Fritzsch for his
hospitality
and encouragements. He is also indebted to the Alexander von Humboldt
Foundation for its financial support.

\newpage

\renewcommand{\baselinestretch}{1.25}

\newpage

%% FOLLOWING LINE CANNOT BE BROKEN BEFORE 80 CHAR
%===================================================figure=========================================
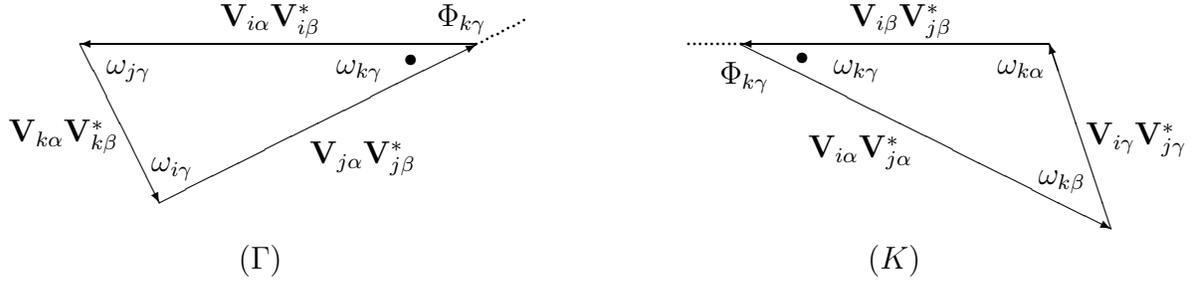
\begin{figure}
\begin{picture}(400,300)
%--------------------------framework----
\put(200,300){\vector(-1,0){150}}
\put(103,307){${\bf V}_{i\alpha}{\bf V}^*_{i\beta}$}
\put(50,300){\vector(1,-2){30}}
\put(23,263){${\bf V}_{k\alpha}{\bf V}^*_{k\beta}$}
\put(80,240){\vector(2,1){120}}
\put(137,255){${\bf V}_{j\alpha}{\bf V}^*_{j\beta}$}
\multiput(200,300)(2,1){10}{\circle*{1}}
\put(185,307){$\Phi_{k\gamma}$}

\put(60,289){$\omega_{j{\gamma}}$}
\put(147,289){$\omega_{k{\gamma}}$}
\put(77,252){$\omega_{i{\gamma}}$}

\put(175,294){\circle*{4}}

\put(110,215){$( \Gamma )$}

%-----------------------------------
\put(416,300){\vector(-1,0){116}}
\put(341,307){${\bf V}_{i\beta}{\bf V}^*_{j\beta}$}
\put(300,300){\vector(2,-1){140}}
\put(325.5,257){${\bf V}_{i\alpha}{\bf V}^*_{j\alpha}$}
\put(440,230){\vector(-1,3){23.5}}
\put(430.5,262){${\bf V}_{i\gamma}{\bf V}^*_{j\gamma}$}
\multiput(300,300)(-2.2,0){10}{\circle*{1}}
\put(292,285){$\Phi_{k\gamma}$}

\put(335,289){$\omega_{k\gamma}$}
\put(397,289){$\omega_{k\alpha}$}
\put(412.5,247){$\omega_{k\beta}$}

\put(323,295){\circle*{4}}

\put(348,215){$( K )$}

\end{picture}
\vspace{-6.5cm}
\caption{The CKM unitarity triangles $\Gamma$ ($=D,S,B$) and $\it K$ ($=U,C,T$)
in
the complex plane. A common outer angle $\Phi_{k\gamma}$ is shared by both
triangles.
Note the relation between the outer angle $\Phi_{k\gamma}$ and the inner angle
$\omega_{k\gamma}$.}
\end{figure}

{}.
\vspace{2.cm}

\begin{center}
\begin{tabular}{l|l}\hline\hline \\
%---------------------------------------------

Promising decay modes 	& Measurable weak phases $\Phi_{i\alpha}$  \\ \\ \hline
\\
%-------------------------
 $\stackrel{(-)}{B}$$^{0}_{d}\rightarrow J/\psi K_{S}$, $D^{+}D^{-}$  	& $\sin
2\Phi_{us}$ \\
 $\stackrel{(-)}{B}$$^{0}_{d}\rightarrow \pi^{+}\pi^{-}$			& $\sin 2\Phi_{cs}$
\\
 $\stackrel{(-)}{B}$$^{0}_{d}\rightarrow \stackrel{(-)}{D}$$^{(*)0}K_{S}$	&
$\sin (\Phi_{us}+\Phi_{cd}-\Phi_{ub})$ \\
 $\stackrel{(-)}{B}$$^{0}_{d}\rightarrow D^{(*)\pm}\pi^{\mp}$		& $\sin
(\Phi_{us}-\Phi_{cs})$ \\ \\

 $\stackrel{(-)}{B}$$^{0}_{s}\rightarrow J/\psi \phi$, $D^{+}_{s}D^{-}_{s}$	&
$\sin 2\Phi_{ud}$ \\
 $\stackrel{(-)}{B}$$^{0}_{s}\rightarrow \rho^{0}K_{S}$				& $\sin
2(\Phi_{cd}-\Phi_{tb})$ \\
 $\stackrel{(-)}{B}$$^{0}_{s}\rightarrow \stackrel{(-)}{D}$$^{(*)0}\phi$		&
$\sin (\Phi_{cd}-\Phi_{ud})$ \\
 $\stackrel{(-)}{B}$$^{0}_{s}\rightarrow \stackrel{(-)}{D}$$^{(*)0}K_{S}$	&
$\sin (\Phi_{ub}-\Phi_{ud}-\Phi_{cs})$ \\  \\ \hline\hline
%-------------------------------------------
\end{tabular}
\end{center}

\vspace{1cm}
\begin{flushleft}
{Table 1:} {Some promising decay modes of $B_{d}$ and $B_{s}$ mesons and
specific rephasing invariant weak phases $\Phi_{i\alpha}$, which are
determinable
in their $CP$ asymmetries.}
\end{flushleft}


\begin{thebibliography}{99}
\bibitem{11} With the Wolfenstein parametrization of the CKM matrix [6],
$J\approx A^{2}\lambda^{6}\eta$.
All current data favor a positive $\eta$, which means $J>0$. The latest
constraints on $\eta$ and
other parameters can be found, e.g., in Ref. [17].
\bibitem{15} Following a different approach, the similar results have been
obtained in Ref. [4].
\end{thebibliography}
\end{document}